# Investigation of the early stage of reactive interdiffusion in the Cu-Al system by in-situ transmission electron microscopy

F. Moisy[a], X. Sauvage[b*], E. Hug[a]

[a] *Normandie Univ, UNICAEN, Ensicaen, CNRS, Laboratoire CRISMAT, 6 Bvd du maréchal Juin 14050 Caen, France*

[b] *Normandie Univ, UNIROUEN, INSA Rouen, CNRS, Groupe de Physique des Matériaux, 76000 Rouen, France*

\* *corresponding author : xavier.sauvage@univ-rouen.fr*

**Abstract**

The early stage of the reactive interdiffusion in the Cu-Al system was investigated at 350°C and 300°C thanks to in-situ transmission electron microscopy. A special care was given to find conditions where the electron beam and the sample free surface do not affect significantly the reaction. A special emphasis was then given on the influence of grain boundaries that are fast diffusion paths, and on nanoscaled particles that may interact with the transformation front. It was found that there is a transient state followed by a steady state where the mean growth rates of intermetallic compounds follow a parabolic law indicating that the kinetics is diffusion controlled. Thanks to the in-situ observations at the nanoscale, it was also possible to track the local velocity of interfaces between the different phases. Strong fluctuations were exhibited within length scales smaller than 100nm and they are partly attributed to interface pinning by nanoscaled particles. Last, considering thermodynamic and kinetic arguments, it is shown that it is mainly an indirect effect induced by a local change of solute fluxes and of concentration gradients.

1. **Introduction**

The design of Cu-Al composites has attracted strong interests since such materials can combine the excellent electrical properties of copper and the lightness and low cost of aluminum [1-8]. Among all the products available on the industrial market, Copper Clad Aluminum (CCA) wires are the most popular. They are produced at the industrial scale and widely used for conductors working with high frequency signals [9-14]. The architecture of CCA wires is relatively simple since they typically exhibit an Al core covered by a Cu skin (typically less than 50µm). The combination of electrical conductivity, mechanical strength and ductility is usually achieved through post-drawing heat treatments at relatively low temperature leading to recovery, partial recrystallization and reactive interdiffusion through Al/Cu interfaces. According to the Cu/Al equilibrium phase diagram [15], a large number of intermetallic compounds (IMC) may appear below 400°C, namely $Al_2Cu$, $AlCu$, $Al_3Cu_4$, $Al_2Cu_3$ and $Al_4Cu_9$. The nucleation and growth of such IMC is of critical importance since they affect both the electrical conductivity and the mechanical behavior. It has been shown indeed that the ultimate tensile stress and the total elongation of CCA wires measured under tensile stress are directly linked to the total thickness of IMC at Al/Cu interfaces [11]. Concomitantly, an increase of the IMC volume fraction leads as well to a significant increase of the electrical resistivity [16-18].

Practically, in annealed CCA wires only three IMC are often observed ($Al_2Cu$, $AlCu$ and $Al_4Cu_9$) [13, 19-21], but it should be noted that in few cases other configurations have been reported involving $Al_3Cu_4$ [7-11], $Al_2Cu_3$ [22], or the absence of $AlCu$ [22, 23]. In any case, the reaction is governed by bulk diffusion and the energy of formation for the growth of the most common IMC are 117 kJ/mol, 107 kJ/mol and 90 kJ/mol for $Al_2Cu$, $Al_4Cu_9$ and $AlCu$, respectively [13]. These values were however determined indirectly from a thermo-kinetic study in the 300-400°C temperature range for annealing times longer than one hour. Thus, these experiments based on Scanning Electron Microscopy (SEM) data do not reflect the early stage of the reaction and do not provide any information about the microscopic details of the transformation such as: i) the influence of defects (like grain boundaries or dislocations) on the kinetics; ii) the influence of particles on the interphase boundary mobility; iii) the sequence of IMC formation (which phase nucleates at first); iv) the growth direction from the original interface. To track such fine scale features, it is necessary to investigate microstructures using Transmission Electron Microscopy (TEM) which also gives the possibility of in-situ annealing experiments to follow the nucleation and growth of the new phases [24].

Based on detailed TEM analyses, Xu and co-authors [25] suggested that the $Al_2Cu$ phase is the first to nucleate at Al/Cu interfaces (in contradiction with thermodynamical analyses indicating lower free energy of formation for the AlCu phase [13]). Then, they observed the nucleation of the $Al_4Cu_9$ phase forming a second layer that concomitantly growths with $Al_2Cu$. They propose that the $Al/Al_2Cu$ interface moves toward the Al side and that the $Al_4Cu_9/Cu$ interface moves toward the Cu side. In a more recent work, the same authors reported in-situ high resolution TEM observations [26] where AlCu seems to appear after the nucleation of $Al_4Cu_9$. However, for all these experiments, samples were made of thin layers (thickness below one micrometer). It necessarily limits the extent of diffusion gradients and probably affects the whole kinetics. In-situ TEM combined with a systematic comparison with ex-situ annealed samples, has been explored on the Al-Cu system by Tan and co-authors [27]. They reported nanoscaled particles (attributed to $Al_3Cu_4$ or $Al_2Cu_3$ phases) inside the thicker layers of IMCs ($Al_2Cu$, AlCu and $Al_4Cu_9$) but only at the early stage of the reaction. They also revealed that oxides located at the Al/Cu interface prior to the reaction were embedded in the $Al_4Cu_9$ phase or cover AlCu / $Al_4Cu_9$ interfaces at the end of the reaction.

These previous studies demonstrate the potentiality offered by in-situ TEM experiments to clarify microscopic mechanisms of the reactive interdiffusion in the Al-Cu system. The aim of the present work was to apply this technique to Al-Cu composites processed by drawing without the intrinsic limitations of thin films and avoiding sample preparation by the focused ion beam (FIB) technique which could introduce defects that may affect the reaction kinetics. A special emphasis has been then given on: 1) the influence of crystalline defects resulting from the plastic deformation (boundaries and dislocations); 2) the impact of oxide particles located at the original Al/Cu interface; 3) the interaction between intragranular particles in the fcc Al matrix and the growing IMC; 4) the real time motion and velocities of the multiple heterophase boundaries that appear during the reactive interdiffusion.

## 2. Experimental procedure

The Al-Cu composite material investigated in the present study was obtained by a specific cold-drawing process that was developed to produce architectured materials [28]. In a first step, Copper (OFHC Cu) and Aluminum (99.5% purity, see table 1) are drawn together to produce a CCA wire, then 60 pieces of this CCA are restacked in a Cu tube and further drawn down to 3mm, leading to a Cu wire that contains 60 embedded continuous Al fibers. The Al-Cu composite material was processed at room temperature to avoid any reaction between Cu and Al. The total level of plastic deformation endured by Cu and Al is estimated by the drawing ratio : $\eta = \text{Ln}\left(\frac{S_0}{S}\right) = 7.4$, with $S_0$ and $S$ the initial and the final section area respectively. Using this procedure, a very large number of Al/Cu interfaces are created, so that TEM samples always exhibit several interfaces in the electron transparent area. It allows the selection of one of them for the observation of the in-situ reaction while others could be investigated after to check any influence of the electron beam on the reactive interdiffusion.

**Table 1 : Nominal compositions (%wt) of initial materials used to process the Al-Cu composite wire**

| 99.5% purity Al rod (Al balanced) | | | | | | |
|---|---|---|---|---|---|---|
| **Si** | **Fe** | **Cu** | **Mg** | **Mn** | **Zn** | **Ti** |
| 0.1 | 0.3 | 0.02 | 0.01 | 0.01 | 0.01 | 0.01 |

| OFHC Cu tubes (Cu balanced) | | | | | | | | | |
|---|---|---|---|---|---|---|---|---|---|
| **Pb** | **S** | **Fe** | **Zn** | **Sn** | **P** | **As** | **Ni** | **Sb** | **Bi** |
| 0.004 | 0.004 | 0.004 | 0.003 | 0.002 | 0.002 | 0.002 | 0.002 | 0.002 | 0.001 |

TEM samples were prepared in the cross section of the Al-Cu composite wire. Discs of 3mm in diameter were sliced, mechanically polished down to 200μm thick and then electropolished using a 70% methanol - 30% nitric acid solution at -30°C with a voltage of 15V. Optimal electron transparency was finally achieved by ion milling using a Gatan precision ion polishing system (PIPS 2).

TEM observations were carried out with a JEOL ARM 200F microscope operated at 200 kV. Bright field (BF), dark-field (DF - collection angles 20 to 80mrad) and high-angle annular dark field (HAADF – collection angles 80 to 300mrad) images were recorded by scanning transmission electron microscopy (STEM) with a probe size of 0.2 nm and a convergence angle of 34 mrad. In-situ TEM experiments were carried out with a double tilt heating holder (Gatan 652 MA) at 300°C and 350°C. In both cases, a stable temperature was reached in less than 2min allowing the observation of the early stage of the interface reaction. Analytical data were also recorded by Energy Dispersive X-ray Spectroscopy (EDS) with a JED2300 detector.

A comparison with samples prepared from the same Al-Cu composite annealed ex-situ at the same temperatures under vacuum ($10^{-6}$ mbar) was carried out to validate the in-situ TEM approach. Due to the brittleness of the interfaces induced by IMC, it was not possible to prepare suitable TEM samples from ex-situ annealed samples using the technique described above. Thus, thin foils were obtained by FIB using a Thermofisher PFIB Hélios. Some additional observations were also carried out by SEM using secondary electrons (SE) and backscattered electrons (BSE) detectors with an acceleration voltage ranging from 10 to 20kV.

## 3. Results

*3.1. Al/Cu interfaces in the as-drawn composite*

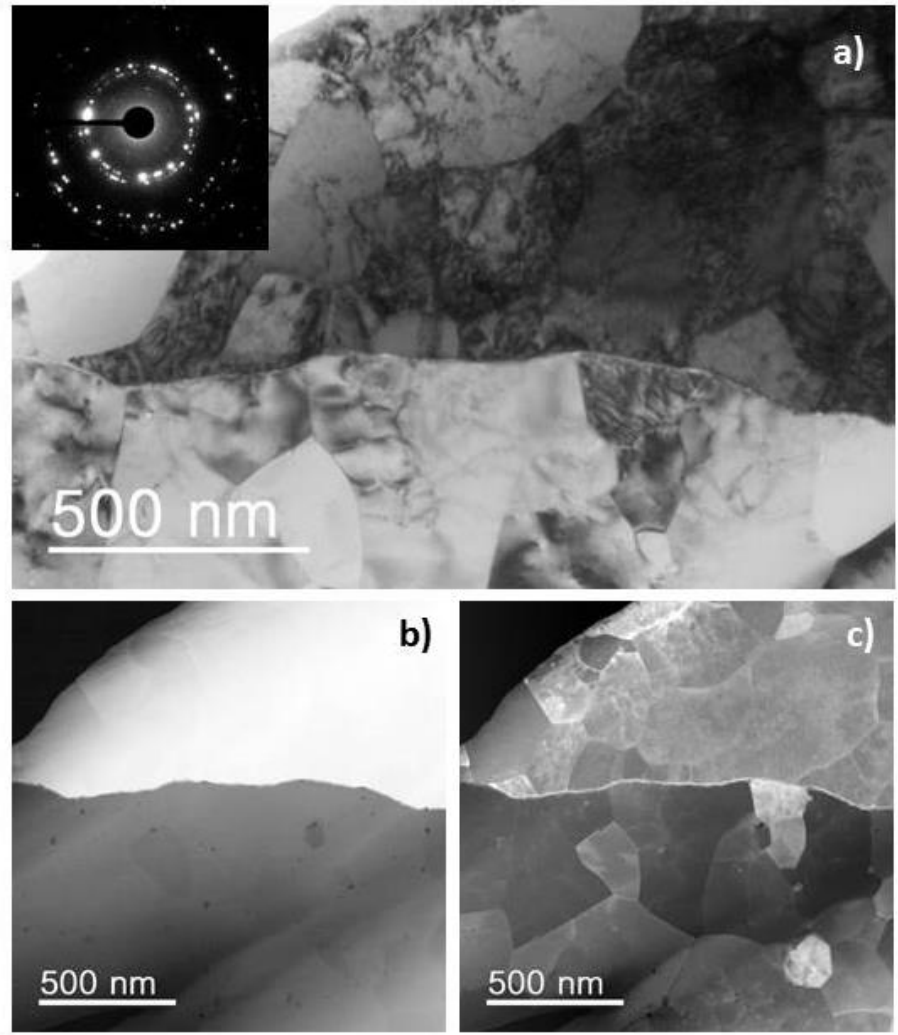

*Figure1: Typical Al/Cu interface in the initial composite wire (Cu top, Al bottom, cross sectional view) ; (a) TEM bright field image showing sub-boundaries resulting from the large level of deformation and corresponding SAED (inset) ; b) STEM-HAADF image where fcc Cu is brightly imaged, some nanoscaled dark particles are also exhibited in the fcc Al ; c) corresponding STEM-DF image.*

A typical Al/Cu interface viewed in the cross section of the initial composite wire is shown in Fig. 1. Thanks to the different atomic number of Cu and Al, the two phases are easily differentiated on STEM-HAADF images (Fig. 1b) where Cu appears brighter. Due to the large level of plastic deformation that was applied during the drawing process, grains were elongated along the wire axis and the mean value of their cross-sectional diameter that appears on the TEM BF image (Fig. 1a) or on the STEM-DF image (Fig. 1c) is in a range of 200 to 300 nm. The Selected Area Electron Diffraction (SAED) pattern (inset in Fig. 1a) exhibits Debbye-Scherrer rings that could be indexed with both fcc Al and fcc Cu phases. It indicates that most of these boundaries are high angle grain boundaries (GB). The contrast appearing inside grains on the TEM BF image (Fig. 1a) indicates significant lattice strains probably due to a high dislocation density resulting from the drawing process. On the STEM-HAADF image, some nanoscaled particles are darkly imaged on the Al side. As shown on the EDS map in Fig. 2, they contain both Si and Mg. These two elements were in the original Aluminum rod used to prepare the composite (table 1), and due to the low solubility of Mg and Si in Al, $Mg_2Si$ particles have nucleated during the annealing treatment carried out before drawing.

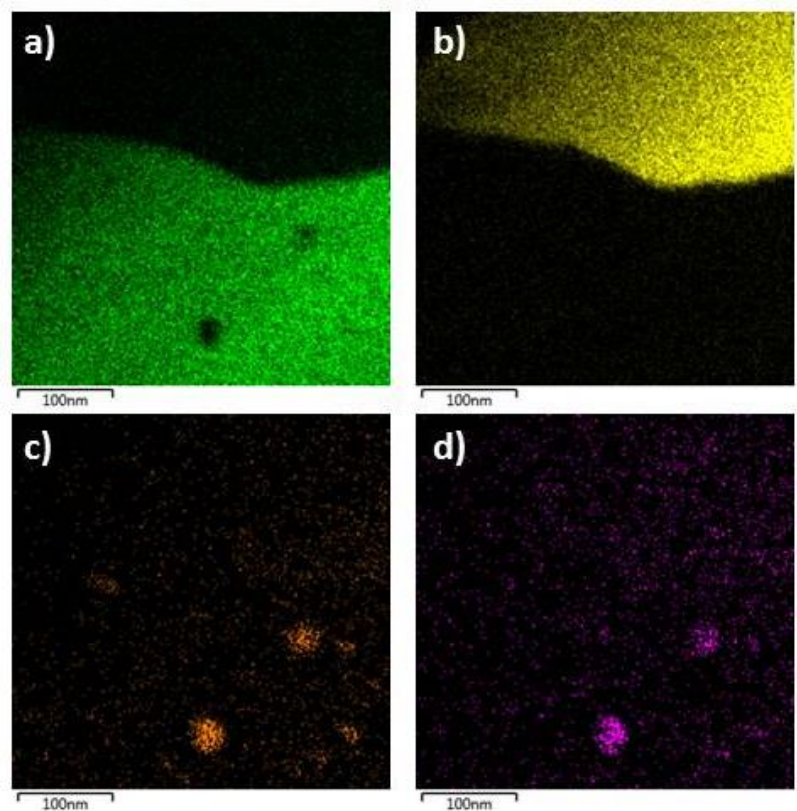

*Figure2: STEM EDS maps ((a) Al, (b) Cu, (c) Mg, (d) Si) showing the Al/Cu interface and nanoscaled $Mg_2Si$ particles in the fcc Al.*

The Al/Cu interface in the as-drawn state of the composite wire is shown in Fig. 3. On the STEM-HAADF image, $Mg_2Si$ particles clearly appear inside Al grains and other nanoscaled particles (less than 10 nm in diameter) with different contrasts are imaged along the Al/Cu interface. Some are darker than Al and others with a grey level standing between that of Al and Cu. To identify these particles, EDS line profile analyses where recorded along arrows noted 1 and 2 in Fig. 3a. They are displayed in Fig. 3c and 3d. A significant level of oxygen is detected in the darker particles (profile 2, Fig. 3d). They are most probably $Al_2O_3$ oxide particles originally located at the aluminum rod surface and captured by the Al/Cu interface during the co-drawing process. Other particles at the interface (profile 1, Fig. 3c) seem to contain a mixture of Al and Cu and thus could be some intermetallic particles that have nucleated during the drawing process. Since they are much smaller that the mean foil thickness (estimated in a range of 50 to 100nm while the mean particle diameter is only about 10nm), it is impossible to determine accurately their composition due to the overlap with the surrounding Al phase. Electron diffraction was used to identify their crystallographic structure, however due to numerous possible overlaps of diffraction spots of Al-Cu IMCs and oxides ($Al_2O_3$ and $CuO_2$) it was not possible to state without any doubt if these particles were AlCu or $Al_2Cu$. The third EDS line profile that is displayed in Fig. 3e was recorded along an Al GB located near the Al/Cu interface (location arrowed on the STEM-DF image Fig. 3b and labeled 3). It clearly shows that copper segregated along the boundary. This could be some copper from the original Al rod used for the processing of the composite (table 1), or some copper that diffused along the GB from the Al/Cu interface during the deformation.

One may assume that these nanoscaled features may affect the reactive interdiffusion kinetics: particles might indeed pin moving interfaces of growing IMC, while crystalline defects (dislocations and GBs) could act as nucleation sites and enhance locally the atomic mobility.

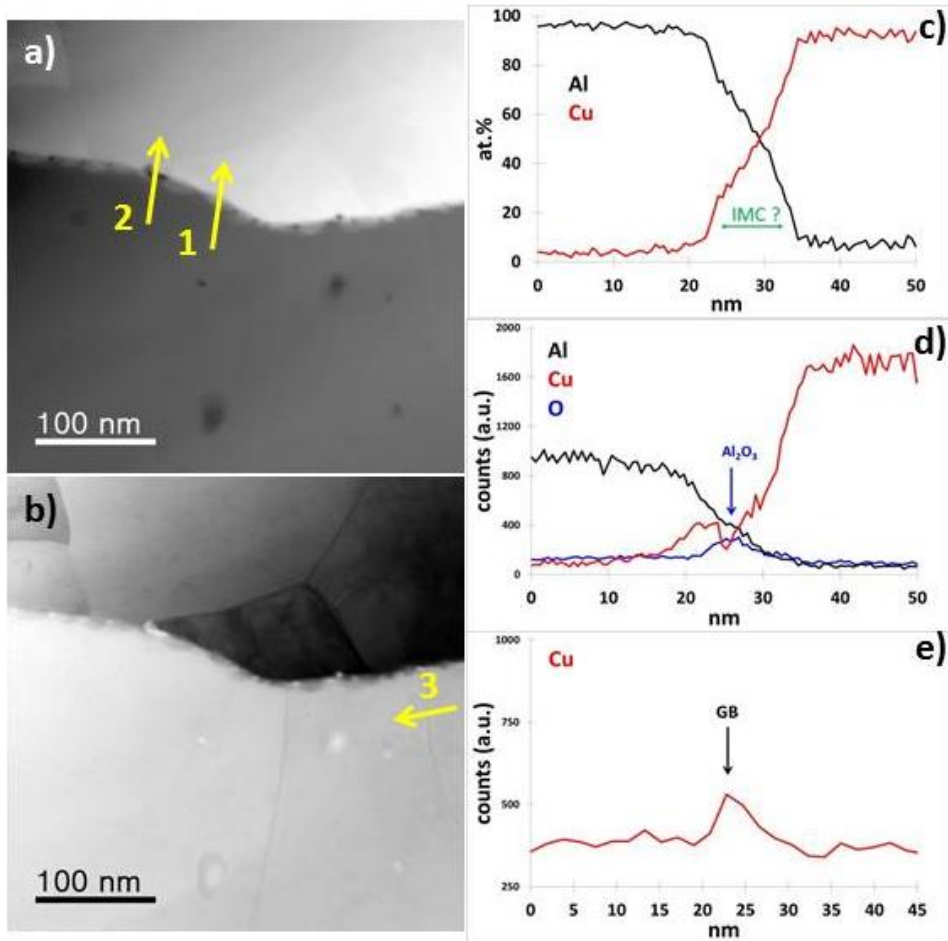

*Figure3: Al/Cu interface in the initial composite material (Al bottom, Cu top) ; a) STEM-HAADF image ; b) STEM-BF image ; c) EDS line profile across the interface (arrow 1 on a)) ; d) EDS line profile across the interface (arrow 2 on a)) ; e) EDS line profile across a grain boundary on the Al side (arrow 3 on b)).*

## 3.2. *Reactive interdiffusion at 350°C observed by in-situ TEM*

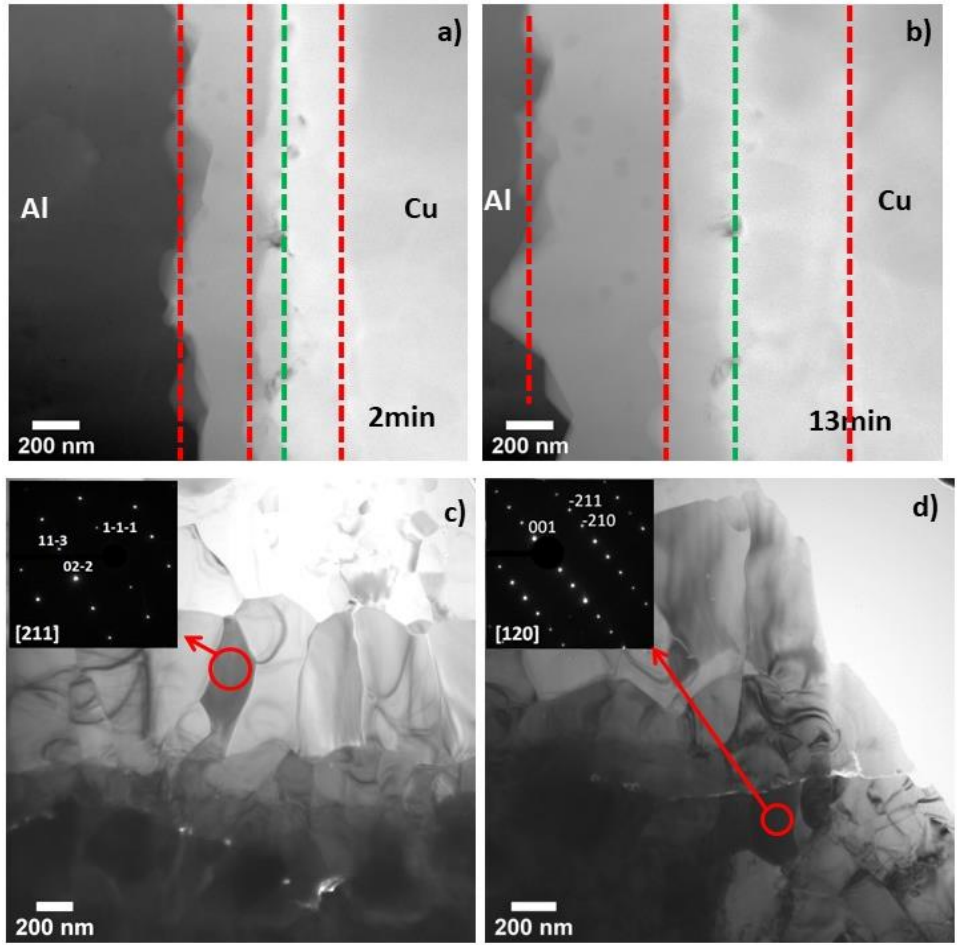

*Figure4: Al/Cu interface after 2min (a) and 13 min (b) at 350°C (STEM-HAADF images where the initial interface is indicated by the green line and interfaces between IMCs and parent phases in red). (c) TEM BF image and SAED pattern recorded in the circled $Al_2Cu$ grain in [211] zone axis. (d) TEM BF image and SAED pattern recorded in the circled $Al_4Cu_9$ grain in [120] zone axis. (c) and (d) are data recorded after the in-situ heating experiments where the sample was held at 350°C during 22min.*

The result of the reactive interdiffusion that occurred at Al/Cu interfaces during the in-situ TEM experiment at 350°C is illustrated in Fig. 4. The IMC layers that form between Al and Cu can be tracked on STEM-HAADF images (Fig. 4a and 4b are snapshot after 2 and 13min respectively) thanks to the Z-contrast. On these images, the original interface is easily identified as it is covered by few $Al_2O_3$ nanoparticles (darkly imaged). For more clarity, this interface is marked by a green line on images. Between Al and Cu, three IMCs layers are clearly exhibited, two of them grew toward Al (left side) and one grew toward Cu (right side). They have different thicknesses and interphase boundaries exhibit a strong roughness at the nanometer scale. At 350°C, the kinetic of the reaction is extremely fast, the total IMC thickness reaches about 500nm in only 2min (Fig. 4a). After 22min held at 350°C, the sample was cooled down to room temperature (50°C was reached in less than a minute) to stop the reaction and to investigate the IMC layer. The Al rich IMC exhibits columnar grains that extend over the whole layer thickness (about 500nm) with a typical width of about 200nm. The crystallographic structure of these grains is consistent with the $Al_2Cu$ phase (Fig. 4c). The Cu rich IMC exhibits more equi-axed grains with a mean size of about 300nm corresponding to the layer thickness. The crystallographic structure of these grains is consistent with the $Al_4Cu_9$ phase (Fig. 4d). The third IMC layer located between $Al_2Cu$ and $Al_4Cu_9$ is the thinnest with also the smallest grains (mean diameter about 200nm). They have been attributed to AlCu. All these observations are in agreement with most of earlier reports about the reactive interdiffusion in the Al-Cu system [12, 13, 16, 20, 28]

There are two possible artefacts that could be introduced during such in-situ TEM experiments, namely the beam effect and the thin foil effect [24]. The electron beam may create defects or a local heating that could both promote the atomic mobility. In the present work, several Al/Cu interfaces were always located in the electron transparency area of the sample but only one was under the electron beam during the in-situ reaction. Thus, after sample cooling it was possible to compare the reaction product at the Al/Cu interface under the electron beam with those out of the beam. After 22min at 350°C, they were all very similar, with the three IMCs having similar thicknesses. We can thus conclude that in our experiments the electron beam has no significant influence on the observed reactions.

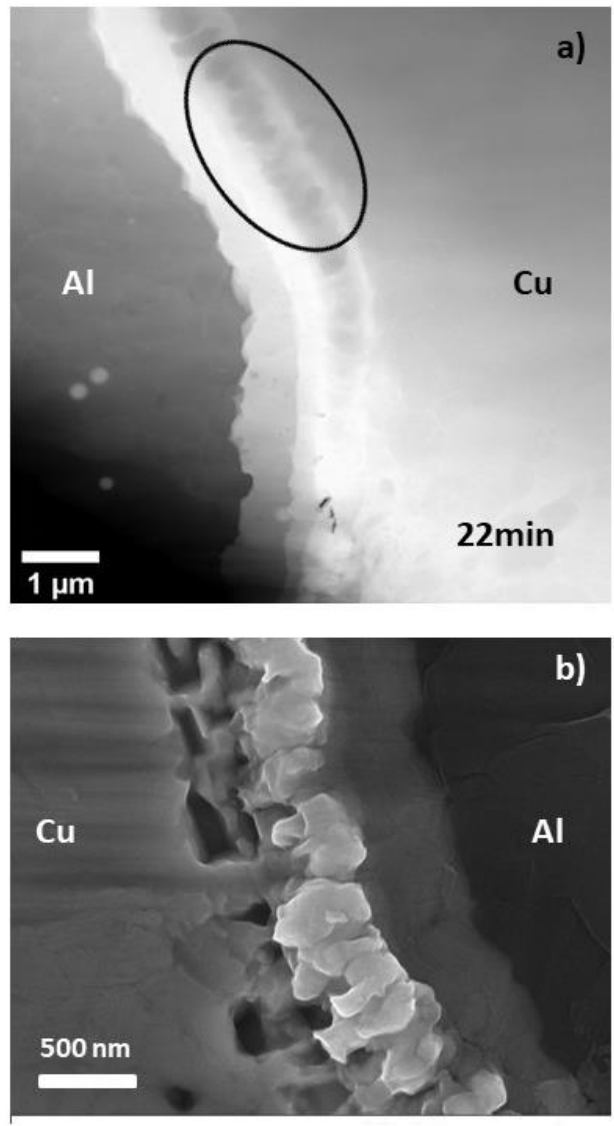

*Figure5: IMCs layer at the Al/Cu interface after 22min at 350°C in-situ in the TEM. (a) STEM-HAADF image showing some contrast inversion in the IMC layer near the fcc Cu phase (circled) ; (b) SEM image showing the morphology of the IMCs layer formed in the thin foil during the in-situ TEM experiment.*

The other possible artefact during in-situ TEM annealing is the thin foil effect which may affect the atomic mobility (through surface diffusion or as a vacancy sink/reservoir) or promote stress relaxation (such stresses may arise from a different molar volume of the nucleated IMCs). Some contrast changes appear on STEM-HAADF images (circled in Fig. 5a) after 22min at 350°C, when the IMC thickness becomes much larger than the foil thickness (about 1 μm against 100nm). As shown on the SEM image taken on the TEM sample after the in-situ heating experiment (Fig. 5b), this is obviously due to a local thickness change resulting from the IMC growth out of the TEM foil surface (the HAADF intensity depends on the mean Z value but also on the local thickness). At this stage of the reaction, they are also numerous nanoscaled cavities located on the Cu side as a result of a Kirkendall effect [29].

*3.3    Reactive interdiffusion at 300°C observed by in-situ TEM*

The reactive interdiffusion at the Al/Cu interface at 300°C gives rise to the same sequence of IMC ($Al_2Cu$, $AlCu$ and $Al_4Cu_9$) with similar growth directions, but as expected, the kinetics is much slower. The STEM-BF image in Fig. 6a shows however that in two minutes only an IMC layer of about 200nm is already formed. Anyway, the slower kinetics allows a detailed investigation of the influence of nanoscaled $Al_2O_3$ lying at the original Al/Cu interface. It seems that the smallest particles do not significantly affect the reaction and the IMC growth at the interface, but particles larger than 50 nm (circled on the top of images in Fig. 6) prevent the nucleation of any IMC on both sides of the interface. Near such particle, the IMCs grow slowly towards the direction perpendicular to the original interface (upward and downward on images in Figs. 6(d), 6(e), 6(f)). After 16min, there is still a small portion of the original interface free of IMC although the mean thickness of IMC is about 400nm (Fig. 6f). Some of $Mg_2Si$ nanoscaled particles located in the fcc Al phase are dissolved due to an increase of the solubility of Mg and Si in Al at 300°C [30] but some remain and also affect the reaction, as they can pin the moving interface boundaries (circled at the bottom of images in Fig. 6). At the end of the process they are fully embedded in the IMC, but they significantly affect the local velocity of boundaries. This point will be addressed in details in the discussion.

Some grain boundaries in aluminum and copper are arrowed in Figs. 6(a). Although they are known as fast diffusion paths [29] and even if some of them exhibit a local solute concentration enrichment (GB segregation, see section 3.1 and Fig. 3e), there is no preferential growth of IMC along GBs and there is no evidence of preferential nucleation along these crystalline defects.

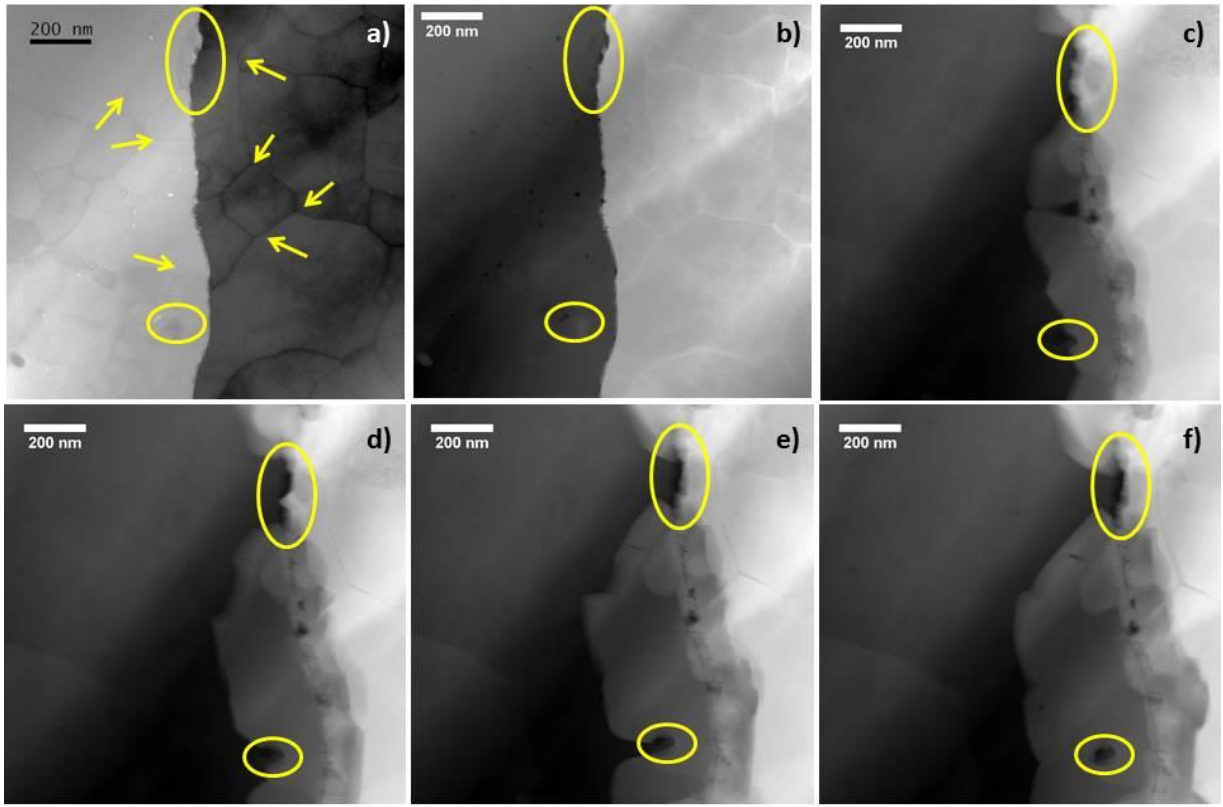

*Figure6: In-situ STEM observations during annealing treatment at 300°C. a) STEM-BF image of the initial interface, b) STEM-HAADF image of the initial interface, c) after 2min at 300°C, (d) 5min, (e) 10min, (f) 16min. Particles that affect the IMC growth are circled (Al2O3 at the interface and Mg2Si in fcc Al). Some of the grain boundaries in Al and Cu are arrowed to show that they do not promote the growth of IMC.*

To evaluate if the kinetics of the reaction could be affected by a thin foil effect during the in-situ TEM observations, the bulk Al-Cu composite material was annealed under vacuum during 30min at 300°C. As shown in Fig. 7, there is no significant visible difference with the sample aged in-situ. IMC layers exhibit a similar roughness, similar compositions (as measured by EDS in Fig. 7b and 7d) and similar thicknesses. Large composition gradients are exhibited in $Al_4Cu_9$ and AlCu phases while it is much more limited in $Al_2Cu$, in agreement with earlier reports [13, 26, 31].

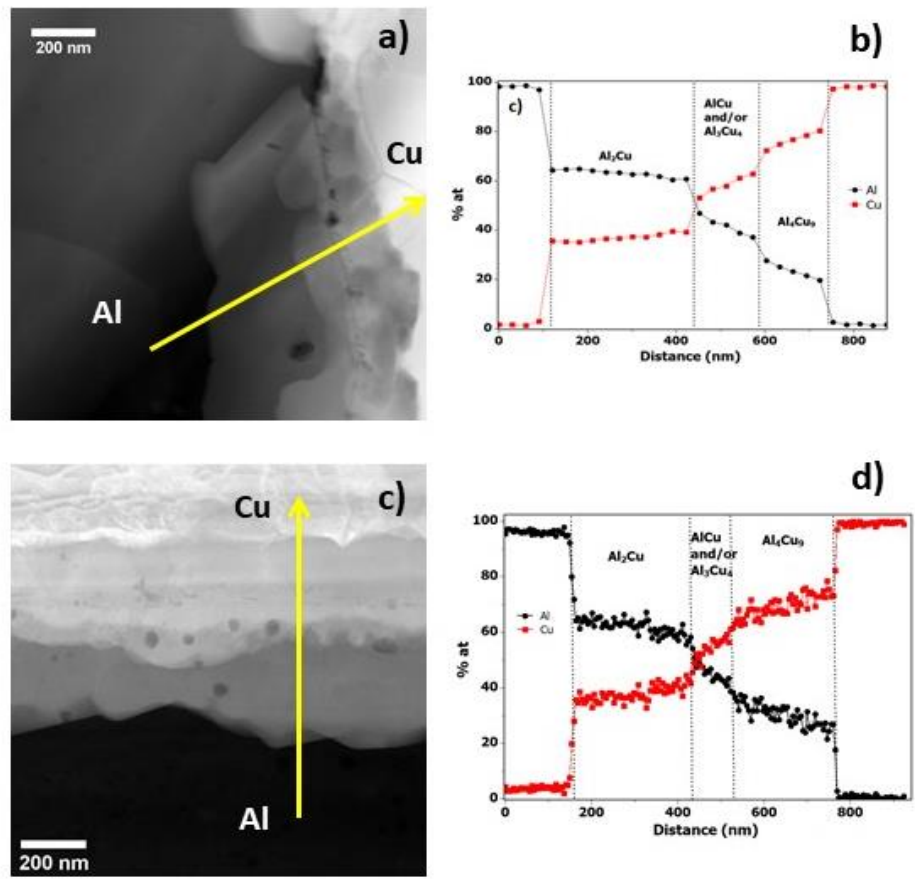

*Figure7: STEM-HAADF images of an Al/Cu interface annealed 30 min at 300°C in situ in the TEM (a) and annealed in similar conditions ex-situ in a bulk sample (c). EDS line profile across the IMC layers (direction arrowed on images) for the in-situ (b) and ex-situ (d) annealed samples.*

## 4. Discussion

In-situ TEM observations performed at 350°C and 300°C allowed studying the interfacial reactions through Al/Cu interfaces leading to the formation of three intermetallic compounds, namely $Al_2Cu$, $AlCu$ and $Al_4Cu_9$. It was not possible to identify unambiguously the first phase that nucleated at the interface because of the extremely fast reaction in the early stage (about 200nm intermetallic layer grown in only two minutes at 300°C (Fig. 6(c))). However, the influence of $Al_2O_3$ oxide particles along initial Al/Cu interfaces was clearly identified. Besides, the growth of each IMC layer can easily be followed. It was shown that it is not affected by GBs in primary Al and Cu phases, but $Mg_2Si$ nanoparticles inside the Al matrix strongly interact with the transformation front. Ex-situ annealing was carried out to check that thin foil or beam effects during in-situ TEM experiments do not significantly affect the reaction.

Previous works on this system have demonstrated that the layer thickness $L_i$ of IMC $i$ follow a parabolic law as a function of time $t$ [29]:

$$L_i^2 = k_p^i \, t \qquad (1)$$

Where $k_p^i$ is the growth rate of IMC $i$.

In case of a growth controlled by the diffusion of one element (A), the balance with flux leads to [29]:

$$\frac{dL_i}{dt} = \frac{D_A \, \Delta_{gi}}{R \, T \, L_i} \qquad (2)$$

where $D_A$ is the intrinsic diffusion coefficient of A (assumed to be not dependent of composition), T the temperature, R the gas constant and $\Delta_{gi}$ the formation energy of IMC i. Then, the combination of eqs. (1) and (2) leads to:

$$k_p^i = \frac{2\,D_A \Delta_{gi}}{RT} \qquad (3)$$

and could also be written as [29]:

$$k_p^i = k_p^{i0} \exp\left(\frac{-Q_i}{RT}\right) \qquad (4)$$

where $k_p^{i0}$ is a constant, $Q_i$ the apparent activation energy of the reaction.

Average IMC layer thicknesses were measured for various annealing times from in-situ TEM data for short annealing times and from SEM images of ex-situ annealed samples for longer annealing times. They are plotted for each IMC as a function of time on the same plot in Figs. 8(a), 8(c), 8(e) and 8(g). A relatively good fit between ex-situ SEM and in-situ TEM data is obtained. The same data were plotted to show the evolution of the square of thicknesses as a function of time (Figs. 8(b), 8(d), 8(f) and 8(h)). The good linearity (regression coefficient $R^2$ > 0.98 excepted for $Al_4Cu_9$ where $R^2 = 0.95$) indicates a parabolic growth with constant $k_p^i$ and thus a diffusion controlled reaction. The growth rate constants estimated from these linear fits are given in table 2. Independently of the temperature, the slowest growing phase is AlCu, the fastest is $Al_2Cu$, and the $Al_4Cu_9$ growth rate lies in-between, in good agreement with data published by other authors on the same system [32, 33].

Table 2: Growth rate constant $k_p^i$ of IMCS estimated from the linear interpolations of plots of Fig. 8

| $k_p^i$ (cm$^2$ / s) | $Al_2Cu$ | AlCu | $Al_4Cu_9$ | All IMC |
|---|---|---|---|---|
| 300°C | 7.7 10$^{-13}$ | 1.8 10$^{-13}$ | 5.5 10$^{-13}$ | 4.2 10$^{-12}$ |
| 350°C | 8.6 10$^{-12}$ | 1.0 10$^{-12}$ | 3.7 10$^{-12}$ | 3.4 10$^{-11}$ |

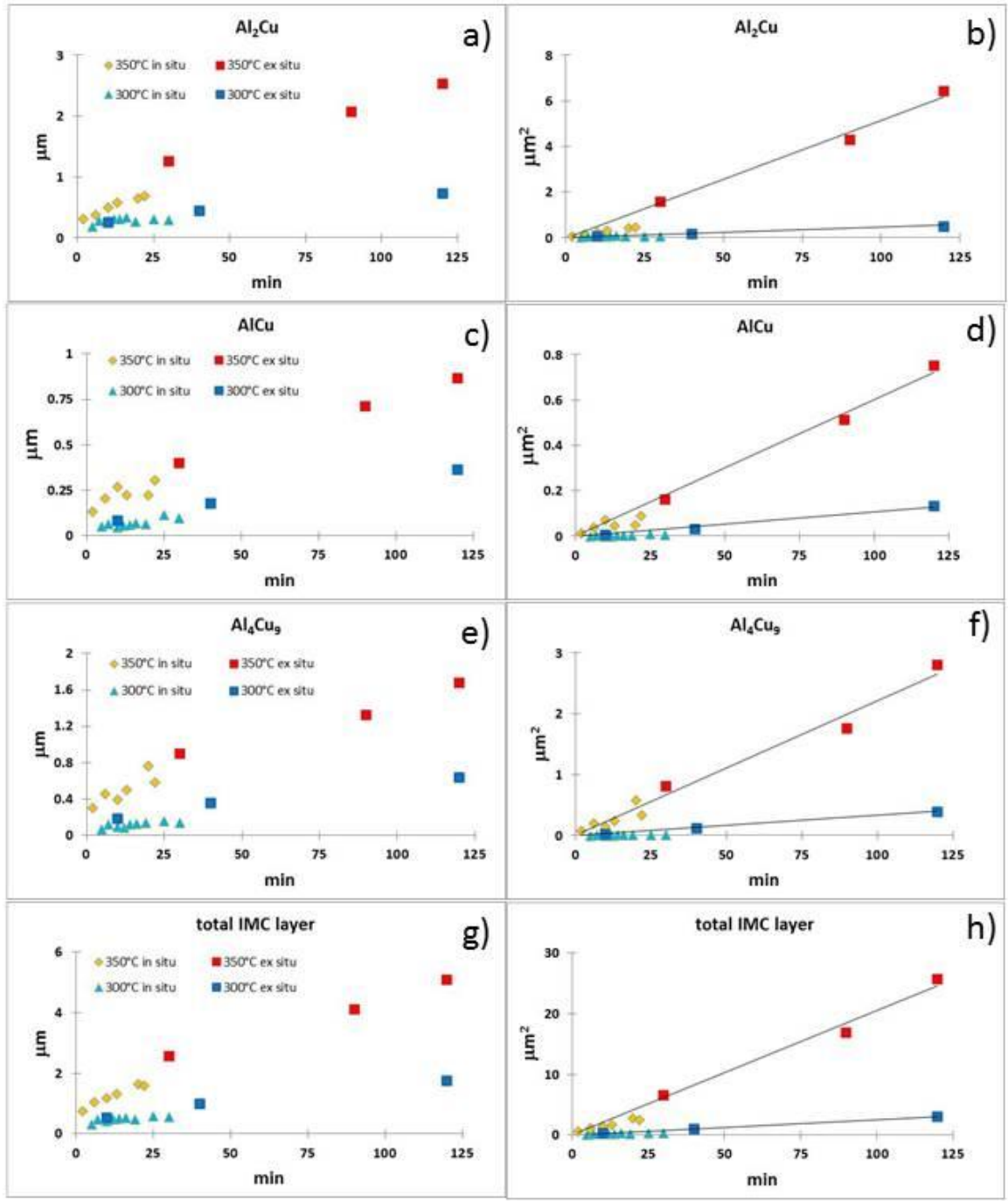

*Figure 8: Thicknesses of Al$_2$Cu (a), AlCu (c), Al$_4$Cu$_9$ (e) and of the full IMC layer (g) as a function of time. Same data plotted as the square of the thickness as a function of time with linear interpolations (Al$_2$Cu (b), AlCu (d), Al$_4$Cu$_9$ (f) and full IMC layer (h)). In-situ TEM and ex-situ SEM data of the reaction at 300°C and 350°C are gathered on the same plots.*

Beyond this general trend, our in-situ TEM observations clearly evidenced that transformation fronts are rough, they interact with nanoparticles (Fig. 6) and their velocity fluctuates. To illustrate this phenomenon, the successive locations of interfaces between phases (Al$_2$Cu/Al, Al$_2$Cu/AlCu, AlCu/Al$_4$Cu$_9$ and Al$_4$Cu$_9$/Cu) are plotted in Fig. 9 (in red, blue, black and yellow respectively). The black line being the original Al/Cu interface, it corresponds to the spatial reference. The Al$_2$Cu phase is growing faster, thus it could be more easily tracked for shorter times. Therefore, they are more red lines than others.

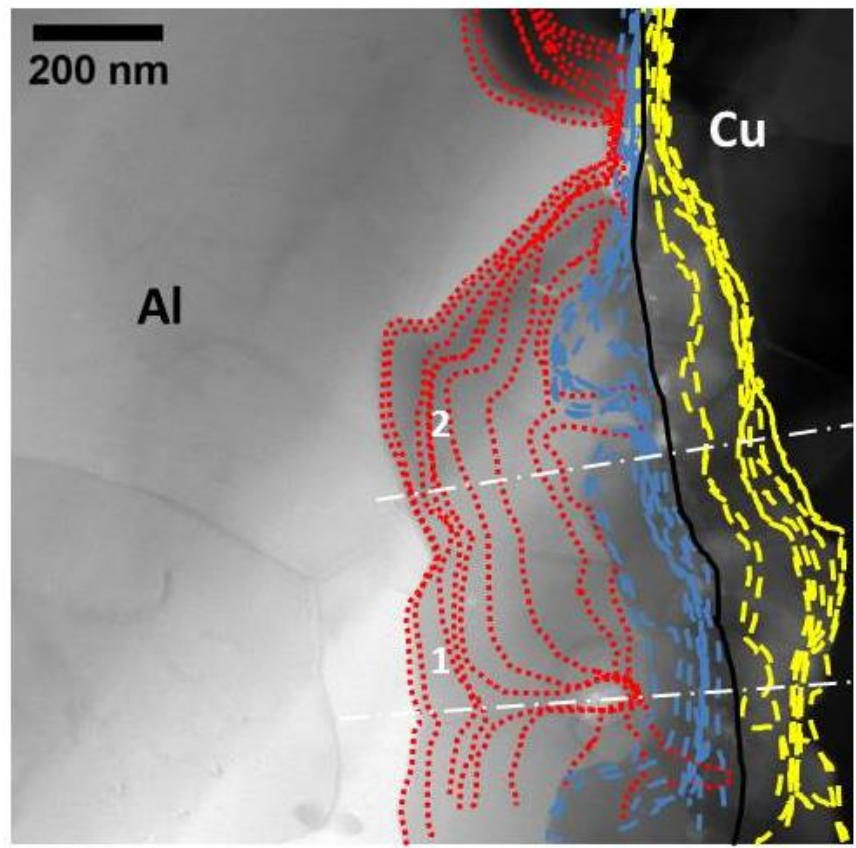

*Figure9: STEM-BF image where the successive positions of the transformation fronts between phases are indicated in dashed lines (corresponding to the sequence displayed in Fig. 6 ). Al/Al$_2$Cu interface in red at t=1, 2, 5, 10, 12, 14, 16, 25, 30min; Al$_2$Cu/AlCu interface in blue at t= 5, 10, 12, 14, 16, 25, 30mi); AlCu/Al$_4$Cu$_9$ in black is the spatial reference; Al$_4$Cu$_9$/Cu interface in yellow at t= 5, 10, 12, 14, 16, 25, 30min. Location 1 was used to monitor the velocity of interfaces in a position where there is interaction with a Mg$_2$Si particle, and location 2 where there is not.*

Two main features are revealed by this figure: i) the distance between two consecutive lines of the same color fluctuates a lot along the interface indicating some large deviations to the average values that could be estimated using eq.(1); ii) there are large differences in the mean position of the transformation front along the interface, especially where a large $Al_2O_3$ particle lies at the original Al/Cu interface. The local IMC thicknesses have been measured on two specific locations to compare velocities where the transformation front meets a $Mg_2Si$ nanoparticle (bottom of image, location 1) and in a place where it is free to move (middle of image, location 2). The corresponding plots are displayed in Fig. 10 with the steady state parabolic law (linear green plot) for comparison. In the middle position (far from the $Mg_2Si$ particle, location labelled 2), at the very beginning of the reaction, the growth of $Al_2Cu$ is much faster than the steady state parabolic law (blue plot in Fig. 10(a)), while other IMCs grow slower (Fig. 10(b) and 10(c)). Then, after about 10 to 15 min, AlCu and $Al_4Cu_9$ exhibit a faster growth rate while that of $Al_2Cu$ significantly decreases. Such feature has already been reported in the literature and it is known that a critical thickness of a given reaction compound might be necessary before others could grow [29]. Besides, the diffusion coefficient of Al in fcc Cu being much slower than the diffusion coefficient of Cu in fcc Al [34], this might also promote the growth of the Al rich intermetallic IMC. In the location labelled 1 on Fig. 9, the growth of $Al_2Cu$ is nearly stopped between 2 and 12min (red plot in Fig. 10(a)) when the transformation front meet the $Mg_2Si$ particle. Then, the particle is overcame, a burst occurs (between 12 and 14 min) and the growth rate reaches that of the middle location (blue line, Fig. 10(a)). In this location, the growth of AlCu is also affected: the growth rate decreases but with a delay of about 10 minutes (between 16 and 25min, see Fig. 10(b)). The growth rate of $Al_4Cu_9$ is only slightly affected, it is a bit reduced and well below the steady state parabolic law (Fig. 10(c)). In average (Fig. 10d), once the particle is passed over by the $Al/Al_2Cu$ interface, both the growth rate and the total thickness of IMCs reach the values of the neighbor region.

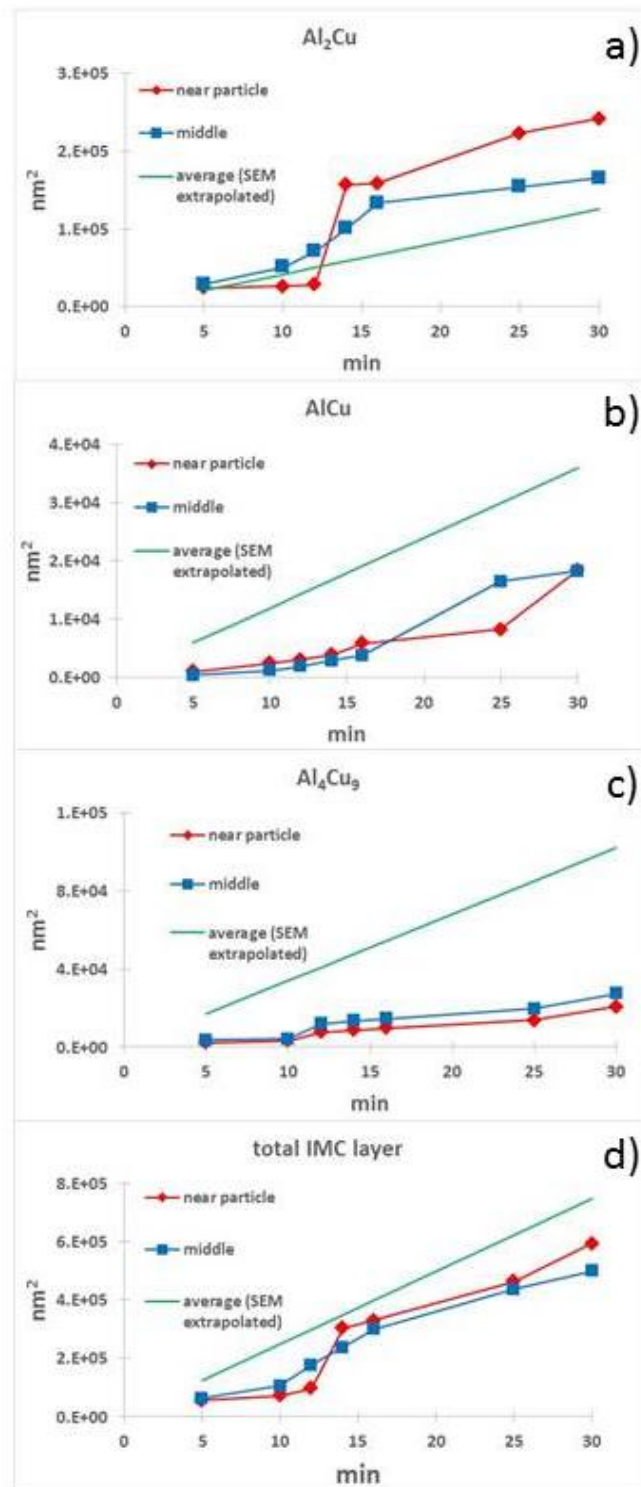

*Figure 10: Square of the thickness of $Al_2Cu$ (a), $AlCu$ (b), $Al_4Cu_9$ (c) and of the full IMC layer (d) as a function of time, measured in two locations on Fig. 8b, near the particle (location 1 indicated on Fig. 8(b)) and in the middle position (location 2 indicated on Fig. 8b). The average value extrapolated from the growth rate measured on SEM images is plotted for comparison.*

The interaction between a particle and a transformation front resulting from reactive interdiffusion could be rather complex [32]. In the present case however, it seems that the $Mg_2Si$ particle simply act as a pining point. Thus, to understand the influence of such a nanoparticle on the mobility of the transformation front, two arguments may be considered, namely a thermodynamic approach and a kinetic analysis (driven by diffusion mechanisms). The transformation front between $Al_2Cu$ and Al is schematically represented in Fig. 11. The driving force to overcome the particle is the transformation of volume V, but to do so, interface S2 should be created and interface S1 should be transformed (from Al/$Mg_2Si$ into Al/$Al_2Cu$). $Mg_2Si$ have specific orientation relationships with the fcc Al matrix [30] driven by a minimization of the interfacial energy, so in the following; this interfacial energy will be considered negligible comparing to Al/$Al_2Cu$ interfacial energy $\gamma_1$. Thus, the energy balance writes as:

$$\Delta G = V \Delta G_{Al_2Cu} + \gamma_1 S_1 + \gamma_2 S_2 \qquad (5)$$

Where $\Delta G_{Al2Cu}$ is the formation enthalpy of $Al_2Cu$, about –100kJ/mol [13]. The transformation could occur only if $\Delta G < 0$.

Assuming that V has a cone shape (as schematically drawn in Fig. 11), then some simple geometric considerations lead to:

$$V = \frac{2 \pi r^3 (1+\tan \alpha)(1+2 \tan \alpha)}{3 (\tan \alpha)^2} \qquad (6)$$

$$S_1 = 2 \pi r^2 \qquad (7)$$

$$S_2 = 4 \pi r^2 \left(\frac{1+\tan \alpha}{\tan \alpha}\right)^2 \qquad (8)$$

With r the radius of the particle and the angle α as defined in Fig. 11. If we assume that $\gamma_1=\gamma_2=1J/m^2$ (upper limit for an incoherent interface in a metallic system) and if $25° < \alpha < 80°$, then combination of eqs. 5, 6, 7, 8 indicates that the transformation will be favorable ($\Delta G < 0$) for any particle bigger than 1nm in diameter. Thus, the reaction near the particle is not delayed because of a high energy barrier, and other arguments have to be considered.

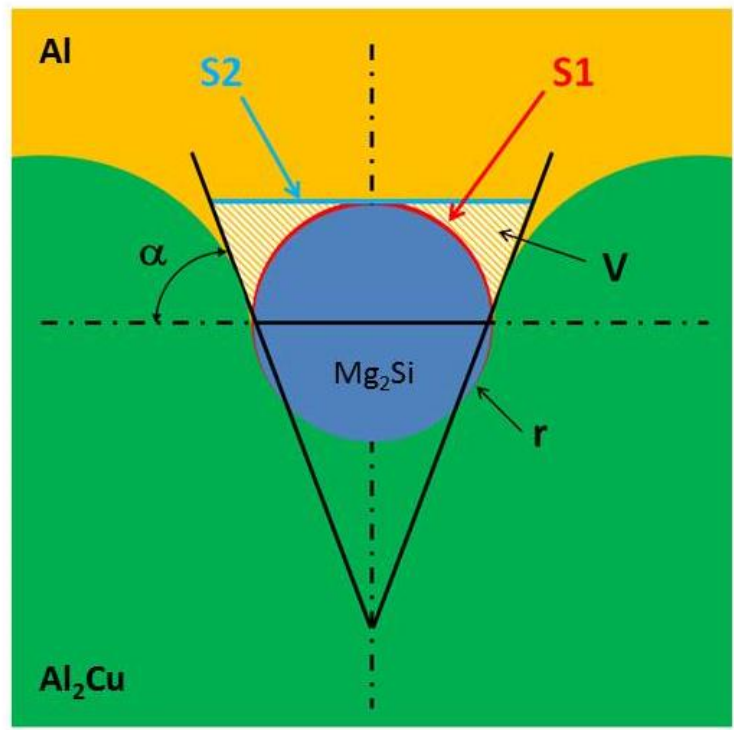

*Figure 11: Schematic representation of the transformation front between Al and Al$_2$Cu interacting with a Mg$_2$Si nanoparticle of radius r. To pass beyond the particle, two interfaces must be created (S1 and S2) which is compensated by the transformation of the volume V (see text for details).*

Equation (3) clearly indicates that a change in $k_p^i$ can only be connected with a change in the diffusion coefficient $D_A$. Of course, the particle cannot affect the intrinsic diffusion coefficient, but in fact the formalism leading to equation (3) is based on atom flux *J* from the Fick's law so that [29]:

$$\vec{J} = -D \, \overrightarrow{grad} C \qquad (9)$$

Where C is solute concentration (in the present case Cu that diffuses in $Al_2Cu$ toward the Al/$Al_2Cu$ interface).

The different steps of the progression of the Al/$Al_2Cu$ transformation front near a nanoscale particle are schematically represented in Fig. 12. Before reaching the particle (step 1), the growth direction is parallel to the diffusion fluxes of Al and Cu atoms in the $Al_2Cu$ phase (respectively the yellow and the red arrows), namely the "y" direction. When it reaches the particle (step 2), since Cu atoms cannot react with Al to grow $Al_2Cu$, then the concentration gradient decreases and fluxes below the contact point O drop while they stay constant in the neighbor region (right side on the image). When the transformation front process further (step 3), some lateral diffusion occurs because a composition gradient develops along the "x" direction (Cu concentration is higher below O than O'). Then the flux near the particle is tilted toward the "y" direction and consequently the growth direction is also tilted (step 4). Then, when the transformation front has already absorbed a large part of the particle (step 5), the contact angle $\alpha$ remains below $\pi/2$ because with a similar flux of solute, a larger surface has to be covered near the particle due to the curvature of the surface. Finally, when the contact angle reaches $\pi/2$ (step 6), lateral composition gradients start to develop above the particle, changing the growing direction and leading at the end to an embedded nanoparticle within the $Al_2Cu$ phase.

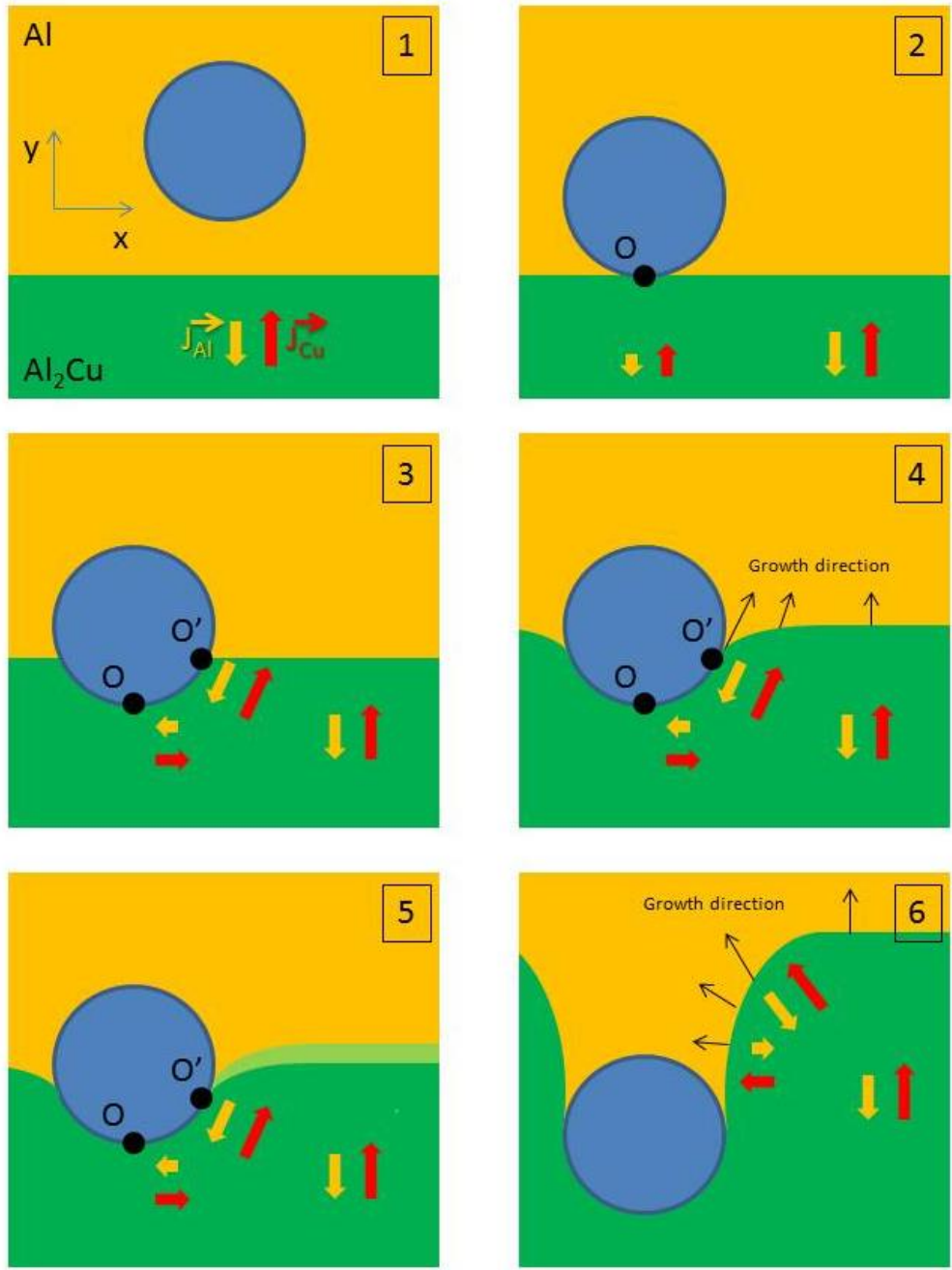

*Figure 12: Schematic representation of the different steps of the interaction between the transformation front between Al and Al$_2$Cu (from 1 to 6). The yellow arrows indicate the Al flux in Al$_2$Cu and the red arrows the flux of Cu atoms (see text for details).*

**Conclusions**

In this work, the early stage of the reactive interdiffusion in the Al/Cu system was investigated by in-situ TEM at 350°C and 300°C. Systematic comparisons with ex-situ annealed samples and with regions out of the electron beam proved that there is no significant artefact (thin foil or beam effect). Original Al/Cu interfaces were created by a purely mechanical process using co-deformation at room temperature by drawing. The initial structure is characterized by a sub-micrometer grain size, few nanoscaled $Al_2O_3$ particles at Al/Cu interfaces and some $Mg_2Si$ particles in the aluminum matrix. As reported by other authors, during the reactive interdiffusion only three IMCs were detected (namely $Al_2Cu$, $AlCu$ and $Al_4Cu_9$) while five are predicted by the equilibrium phase diagram. $Al_2Cu$ and $AlCu$ grow in the Al side and the $Al_4Cu_9$ in the Cu side. Although GBs may act as fast diffusion path, there is no preferential growth of IMC along these defects located both in the original Al and Cu phases. The mean growth rates of all IMCs follow a classical parabolic law with rate constant in agreement with earlier reports. It indicates that the kinetic of the transformation is controlled by diffusion mechanisms. A strong deviation was observed however in the early stage of the reaction where the growth rate of $Al_2Cu$ is significantly larger and those of $AlCu$ and $Al_4Cu_9$ smaller than in the steady state. Oxide particles located at the Al/Cu interface do not inhibit the reaction, unless they cover a large surface (larger than 100x100 nm), while nanoscaled $Mg_2Si$ particles located in Al exhibit a strong interaction with the transformation front. It leads to large fluctuations of the velocity of interphase boundaries within a length scale as small as a 100 nm. Using thermodynamics arguments, it was shown that it cannot account for the pinning effect of nanoscaled particles on $Al/Al_2Cu$ interfaces. Indeed, a simple analytical calculation showed that passing these obstacles is largely favorable from a thermodynamic point of view. Then, we conclude that particles indirectly pin interfaces because they affect the local flux of Cu and Al. It affects composition gradients and consequently the local growth direction of $Al_2Cu$ at the interface with Al.

**Acknowledgments**

This work was supported by "Institut Carnot ESP" fundings, CONDULIGHT Project. Authors would like to thank Nga Nguyen for the wire drawing and F. Cuvilly for the TEM sample preparation by FIB. TEM experiments have been carried out on the GENESIS facility which is supported by the Région Normandie, the Métropole Rouen Normandie, the CNRS via LABEX EMC3 and the French National Research Agency as a part of the program "Investissements d'avenir" with the reference ANR-11-EQPX-0020.

**References**

[1] A. Abdollah-Zadeh, T. Saeid, B. Sazgari, Microstructural and mechanical properties of friction stir welded aluminum/copper lap joints, J. Alloys Compd. 460 (2008) 535–538.

[2] L.Y. Sheng, F. Yang, T.F. Xi, C. Lai, H.Q. Ye, Influence of heat treatment on interface of Cu/Al bimetal composite fabricated by cold rolling, Compos. Part B Eng. 42 (2011) 1468–1473.

[3] M.R. Toroghinejad, R. Jamaati, J. Dutkiewicz, J.A. Szpunar, Investigation of nanostructured aluminum/copper composite produced by accumulative roll bonding and folding process, Mater. Des. 51 (2013) 274–279.

[4] R. Lapovok, H.P. Ng, D. Tomus, Y. Estrin, Bimetallic copper–aluminium tube by severe plastic deformation, Scr. Mater. 66 (2012) 1081–1084.

[5] H. Huang, Y. Dong, M. Yan, F. Du, Evolution of bonding interface in solid–liquid cast-rolling bonding of Cu/Al clad strip, Trans. Nonferrous Met. Soc. China. 27 (2017) 1019–1025.

[6] M. Eizadjou, A. Kazemitalachi, H. Daneshmanesh, H. Shakurshahabi, K. Janghorban, Investigation of structure and mechanical properties of multi-layered Al/Cu composite produced by accumulative roll bonding (ARB) process, Compos. Sci. Technol. 68 (2008) 2003–2009.

[7] A.E. Medvedev, R. Lapovok, E. Koch, H.W. Höppel, M. Göken, Optimisation of interface formation by shear inclination: Example of aluminium-copper hybrid produced by ECAP with back-pressure, Mater. Des. 146 (2018) 142–151.

[8]  T. Jin, G. Li, Y. Cao, R. Xu, S. Shao, B. Yang, Experimental research on applying the copper-clad aluminum tube as connecting tubes of air conditioners, Energy Build. 97 (2015) 1–5.

[9]  N. Guan, C. Kamidaki, T. Shinmoto, K. 'ichiro Yashiro, AC Resistance of Copper Clad Aluminum Wires, IEICE Trans. Commun. E96.B (2013) 2462–2468.

[10]  C.R. Sullivan, Aluminum Windings and Other Strategies for High-Frequency Magnetics Design in an Era of High Copper and Energy Costs, in: IEEE, 2007: pp. 78–84.

[11]  E. Hug, N. Bellido, Brittleness study of intermetallic (Cu, Al) layers in copper-clad aluminium thin wires, Mater. Sci. Eng. A. 528 (2011) 7103–7106.

[12]  A. Gueydan, E. Hug, Secondary creep stage behavior of copper-clad aluminum thin wires submitted to a moderate temperature level, Mater. Sci. Eng. A. 709 (2018) 134–138.

[13]  A. Gueydan, B. Domengès, E. Hug, Study of the intermetallic growth in copper-clad aluminum wires after thermal aging, Intermetallics. 50 (2014) 34–42

[14]  T.T. Sasaki, R.A. Morris, G.B. Thompson, Y. Syarif, D. Fox, Formation of ultra-fine copper grains in copper-clad aluminum wire, Scr. Mater. 63 (2010) 488–491.

[15]  O. Zobac, A. Kroupa, A. Zemanova, K.W. Richter, Experimental description of the Al-Cu binary phase diagram, Metall and Mat. Trans A 50 (2019) 3805-3815

[16]  J. Zhang, B. Wang, G. Chen, R. Wang, C. Miao, Z. Zheng, W. Tang, Formation and growth of Cu–Al IMCs and their effect on electrical property of electroplated Cu/Al laminar composites, Trans. Nonferrous Met. Soc. China. 26 (2016) 3283–3291.

[17]  J.A. Rayne, M.P. Shearer, C.L. Bauer, Investigation of interfacial reactions in thin film couples of aluminum and copper by measurement of low temperature contact resistance, Thin Solid Films. 65 (1980) 381–391.

[18]  W.-B. Lee, K.-S. Bang, S.-B. Jung, Effects of intermetallic compound on the electrical and mechanical properties of friction welded Cu/Al bimetallic joints during annealing, J. Alloys Compd. 390 (2005) 212–219.

[19]  S. Tavassoli, M. Abbasi, R. Tahavvori, Controlling of IMCs layers formation sequence, bond strength and electrical resistance in AlCu bimetal compound casting process, Mater. Des. 108 (2016) 343–353.

[20]  D. Moreno, J. Garrett, J.D. Embury, A technique for rapid characterization of intermetallics and interfaces, Intermetallics. 7 (1999) 1001–1009.

[21]  M. Abbasi, A. Karimi Taheri, M.T. Salehi, Growth rate of intermetallic compounds in Al/Cu bimetal produced by cold roll welding process, J. Alloys Compd. 319 (2001) 233–241.

[22] V.N. Danilenko, S.N. Sergeev, J.A. Baimova, G.F. Korznikova, K.S. Nazarov, R.K. Khisamov, A.M. Glezer, R.R. Mulyukov, An approach for fabrication of Al-Cu composite by high pressure torsion, Mater. Lett. 236 (2019) 51–55.

[23] H. Xu, C. Liu, V.V. Silberschmidt, Z. Chen, Growth of Intermetallic Compounds in Thermosonic Copper Wire Bonding on Aluminum Metallization, J. Electron. Mater. 39 (2010) 124–131.

[24] Q. Dong, P. Saidi, H. Yu, Z. Yao, M.R. Daymond, A direct comparison of annealing in TEM thin foils and bulk material: application to Zr-2.5Nb-0.5Cu alloy, Mat. Charact. 151 (2019) 175-181

[25] H. Xu, C. Liu, V.V. Silberschmidt, S.S. Pramana, T.J. White, Z. Chen, V.L. Acoff, Behavior of aluminum oxide, intermetallics and voids in Cu–Al wire bonds, Acta Mater. 59 (2011) 5661–5673.

[26] H. Xu, I. Qin, H. Clauberg, B. Chylak, V. L. Acoff, New observation of nanoscale interfacial evolution in micro Cu-Al wire bonds by in-situ high resolution TEM study, Scripta Mater 115 (2016) 1-5.

[27] Y.Y. Tan, Q.L. Yang, K.S. Sim, L.T. Sun, X. Wu, Cu–Al intermetallic compound investigation using ex-situ post annealing and in-situ annealing, Microelectron. Reliab. 55 (2015) 2316–2323.

[28] F. Moisy, A. Gueydan, X. Sauvage, A. Guillet, C. Keller, E. Guilmeau, E. Hug, Influence of intermetallic compounds on the electrical resistivity of architectured copper clad aluminum composites elaborated by a restacking drawing method, Mater. Des. 155 (2018) 366–374.

[29] A. Paul, T. Laurila, V. Vuorinen, S. V. Divinski, Thermodynamics, Diffusion and the kirkendall effect in solids, Springer Cham Heildelberg New York Dordrecht London, 2014

[30] S.J. Andersen, C.D. Marioara, A. Frøseth, R. Vissers, H.W. Zandbergen, Crystal structure of the orthorhombic U2-Al4Mg4Si4 precipitate in the Al–Mg–Si alloy system and its relation to the β' and β phases, Materials Science and Engineering A 390 (2005) 127–138

[31] J.M. Sientins, J.W. Gillepsie, S.G. Advani, Transmission electron microscopy of an ultrasonically consolidated copper–aluminum interface, J. Mater. Res. 29 (2014) 1970-1977.

[32] Peng Li, Longwei Pan, Xiaohu Hao, Shuai Li and Honggang Dong, Effect of post-weld heat treatment on inhomohgeneity of aluminum/copper rotary friction welded joint, Mat. Res. Express 5 (2018) 096504

[33] Yajie Guo, Guiwu Liu, Haiyun Jin, Zhongqi Shi, Guanjun Qiao, Intermetallic phase formation in diffusion-bonded Cu/Al laminates, J. Mater Sci 46 (2011) 2467-2473

[34] Dandan Liu, Lijun Zhang, Yong Du, Honghui Xu, Shuhong Liu, Libin Liu, Assessment of atomic mobilities of Al and Cu in fcc Al-Cu alloys, CALPHAD: Computer Coupling of Phase Diagrams and Thermochemistry 33 (2009) 761-768

[35] Anthony De Luca, Michaël Texier, Alain Portavoce, Nelly Burle, Catherine Grosjean, Stéphane Morata and Fabrice Michel, Mechanism of β-Fe-Si$_2$ precipitates growth-and-dissolution and pyramidal defects' formation during oxidation of Fe-contaminated silicon wafers, Journal of Applied Physics 117 (2015) 115302